\documentstyle[epsf,12pt]{article}

\def\bc{\begin{center}}
\def\ec{\end{center}}
\def\be{\begin{equation}}
\def\ee{\end{equation}}
\newcommand{\beqn}{\begin{eqnarray}}
\newcommand{\eeqn}{\end{eqnarray}}

\begin{document}
\pagestyle{empty} 
\vspace{-0.6in}
\begin{flushright}
ROME prep. 97/1164 \\
SNS/PH/1997-005\\
SWAT/148
\end{flushright}
\vskip 0.2 cm
\centerline{\LARGE{\bf{A High Statistics Lattice Calculation}}}
\centerline{\LARGE{\bf{of Heavy-Light Meson Decay Constants}}}
\vskip 1.4cm
\centerline{\bf{C.R.~Allton$^1$, L.~Conti$^{2}$, M.~Crisafulli$^2$,
L.~Giusti$^{3}$, G.~Martinelli$^{2}$, F.~Rapuano$^2$}}
\centerline{$^1$ Dep. of Physics, University of Wales Swansea, Singleton
Park,}
\centerline{Swansea, SA2 8PP, United Kingdom.}
\centerline{$^2$ Dip. di Fisica, Univ. ``La Sapienza"  and}
\centerline{INFN, Sezione di Roma, P.le A. Moro, I-00185 Rome, Italy.}
\centerline{$^3$ Scuola Normale Superiore, P.zza dei Cavalieri 7 and}
\centerline{INFN, Sezione di Pisa, 56100 Pisa, Italy.}
\abstract{We present a high statistics study
of the $D$- and $B$-meson decay constants. The results were obtained
by using the Clover and Wilson lattice actions
at two different values of the lattice spacing $a$, corresponding
to $\beta=6.0$ and $6.2$. After a careful analysis of the
systematic errors present in the extraction of the physical results,
by assuming quite conservative discretization errors,  we find
$f_{D_s}=237 \pm 16 $~MeV, 
$f_{D} = 221 \pm 17 $~MeV ($f_{D_s}/f_D=1.07(4)$), 
$f_{B_s} = 205 \pm 35 $~MeV,
$f_{B} = 180 \pm 32 $~MeV ($f_{B_s}/f_B=1.14(8)$),
in good agreement with previous estimates.}
\vfill\eject
\pagestyle{empty}\clearpage
\setcounter{page}{1}
\pagestyle{plain}
\newpage 
\pagestyle{plain} \setcounter{page}{1}
 
\section{Introduction} \label{intro}
 $f_B$  is a relevant  parameter in phenomenological studies of the 
Standard Model, in determinations of the elements of the 
Cabibbo--Kobayashi--Maskawa (CKM) 
matrix and in studies of $B$--$\overline{B}$ mixing. 
\par The leptonic decay width of the $B$ meson is given by
\be 
\Gamma(B^+ \to \tau^+ \nu_\tau)=\frac{G_F^2 \vert V_{ub}\vert^2}{8 \pi}
M_B \left( 1 - \frac{M_\tau^2}{M_B^2}\right)^2 M_\tau^2 f_B^2 \label{eq:lepto}
\ .
\ee
A precise knowledge of the leptonic decay constant $f_B$, analogous to $f_\pi$
in $\pi \to \mu \nu_\mu$ decays~\footnote{ We use the normalization convention
in which $f_\pi=132$~MeV.}, would then allow an accurate extraction of the CKM 
matrix
element $\vert V_{ub}\vert$, which is actually known with
a relative error of about $25 \%$.\par 
 $f_B$ also enters in  phenomenological analyses 
of the $B^0$--$\bar B^0$ mixing amplitudes, together with the so-called
renormalization group invariant $B$-parameter $\hat B_B$. The 
square of the mixing parameter  $\xi = f_B\sqrt{ \hat B_B}$
is in fact related to the matrix element of the renormalized
$\Delta B=2$  Hamiltonian~\cite{burasdb2}.  All theoretical
calculations of
 $\hat B_B$ tend to give values very close to one for both the $B^0_d$
and the $B^0_s$ mesons. Thus the strength of the mixing is essentially
regulated by the meson decay constant.  It was realized a long time ago
\cite{reina} that a
value of $\xi  \sim 200$~MeV, combined with a large value of the top mass,
leads to a  large value  of $\sin 2\beta$, the parameter
which  controls CP violation in $B \to J/\psi K_s$ decays
\be 
{\cal A} =\frac{N(B^0_d \to J/\psi K_s)(t)- N(\bar B^0_d \to J/\psi K_s)
(t)}{N(B^0_d \to J/\psi K_s)(t)+ N(\bar B^0_d \to J/\psi K_s)
(t)}=  \sin 2 \beta \sin \Delta M_{B_d} t \ . 
\ee
Thus a precise theoretical determination of $f_B$ is very important. \par 
 One of the most
precise determination of $f_B$ is presently given by lattice QCD calculations.
Leptonic decay constants are usually computed from the  matrix element of the
lattice axial current:
\be	Z_A \langle 0|A_0 |P(\vec p=0)\rangle =i f_{P} M_P, \label{fp} \ee
where $M_P$ is the mass of the pseudoscalar meson,
$A_0= \bar Q(x) \gamma_0 \gamma_5 q(x)$, where $Q$ and $q$ denote the heavy
and light quark fields respectively, is
the fourth component of the lattice (local) axial current, and
$Z_A$ is the renormalization constant necessary to relate the lattice
operator to the continuum one \cite{ks,boc}.
In the case of $f_{\pi}$, $f_K$, and more recently $f_{D_s}$, this
is an example of a simple quantity for which a comparison of lattice 
results with experimental data is possible (we notice {\it en passant} that
$f_{D_s}$ was predicted by lattice calculations long before its measurement
\cite{gavela}). \par
\begin{table}
\begin{center}
\begin{tabular}{||c|ccccc||}
\hline\hline         %123456
Run&C60&C62&W60&W62a&W62b\\
\hline
$\beta$&$6.0$&$6.2$ &$6.0$&$6.2$&$6.2$\\
Action & SW & SW & Wil & Wil& Wil\\
\# Confs&170&250&120&250&110\\
Volume&$18^3\times 64$&$24^3\times 64$&$18^3\times 64$ &$24^3\times 64$ 
&$24^3\times 64$\\
\hline
$K_l$&  -   &0.14144&  -    &0.1510&  -   \\
     &0.1425&0.14184&0.1530 &0.1515&0.1510\\
     &0.1432&0.14224&0.1540 &0.1520&0.1520\\
     &0.1440&0.14264&0.1550 &0.1526&0.1526\\
\hline
$K_H$&0.1150&0.1210&0.1255 &0.1300&0.1300\\
     &0.1200&0.1250&0.1320 &0.1350&0.1350\\
     &0.1250&0.1290&0.1385 &0.1400&0.1400\\
     &0.1330&0.1330&0.1420 &0.1450&0.1450\\
     &   -  &  -   &0.1455 &  -  &0.1500\\
\hline
$t_1$-$t_2$&15-28&20-28&15-28&20-28&20-28\\\hline
$a^{-1}(\sigma)$\cite{stringa}&1.88(2)&2.55(1)&1.88(2)&2.55(1)&2.55(1) \\
$a^{-1}(M_\rho)$&1.92(11)&2.56(21)&2.19(9)&2.88(14)&2.92(20)\\
$a^{-1}(f_\pi)$&2.00(6)&2.73(18)&1.91(6)&2.95(14)&2.96(11) \\ 
$a^{-1}(K^*)$-lp-plane\cite{aggr}&2.00(10)&2.7(1)&2.25(6)&3.0(1)&3.0(1) \\
\hline\hline
\end{tabular}
\end{center}
\caption{\it{Summary of the parameters of the runs analyzed in this work.}}
\label{tab:parameters} 
\end{table}
The major sources of uncertainty in the determination of $f_P$, 
besides the effects due to the use of the quenched approximation, come
from the calculation of the constant $Z_A$ in eq.~(\ref{fp}) and
from  discretization errors of $O(a)$, $a$ being the lattice spacing, present
in the  operator matrix elements. The use of chiral Ward identities
allows a non-perturbative determination of $Z_A$
\cite{boc,mm},  thus eliminating
this  source of error. Errors of order $a$ can be reduced by the use
of  ``improved"  lattice actions \cite{sw,luescher}.
Another method to get rid of $Z_A$ consists of extracting 
the decay constants of  heavier pseudoscalar
mesons by computing the ratio $R_P = f_P / f_{\pi}$
and multiplying $R_P$ by the experimental value of the pion decay constant.
Hopefully, by taking the ratio, some of the  $O(a)$ effects are eliminated.
 These  effects  are expected to be more important for $f_D$ than for 
$f_{\pi,K}$
 since, at current values of $a$, the relevant parameter $m_{{\rm charm}} a $ 
 is not very small.  
\par In the last few years, a wide set of results  for $f_P$, obtained 
in the quenched approximation by using
several lattice quark actions and from numerical simulations
on different lattice volumes and at different values of the lattice spacing,
have appeared in the literature \cite{flynn}. Some of these studies
tried to extrapolate the decay constant to the continuum limit
$a=0$  by using values of $f_P$ obtained at different values of the
lattice spacing. If successful, this extrapolation would eliminate
the error due to discretization. The
extrapolation, however,  is difficult because either one has to use
results obtained at large values of $a$, where the dynamics is very different
from that of the continuum limit,
or   the range of $a$ is too small  to observe and correct the $O(a)$ effects 
with a sufficient precision for the extrapolation. For these reasons,
the conclusions of these studies are not very convincing and have changed 
in time.
\par In this paper, we present the results of  a high statistics study
of $f_P$,  at two different values of the lattice spacing,
corresponding to $\beta=6.0$ and $\beta=6.2$, obtained with
the Wilson and the SW-Clover actions \cite{sw}. 
We have preferred to concentrate
our computational  effort on two values of $\beta$ which have been chosen:
\par a) small enough to obtain 
accurate results on  reasonably large physical volumes;
\par b) large enough to avoid the dangerous strong coupling region, which
sits at around $\beta=5.7$. 
\par 
The main parameters of the numerical simulations are given
in table \ref{tab:parameters}. A comparison of the results from two different
actions allows us to study the reduction of discretization errors
in the improved case and to verify the validity of some recipes that 
have been proposed to correct $O(a)$ effects in the Wilson case \cite{lm,klm}
\footnote{ A study of the same problem, performed
  by using the  Ward identities to determine the renormalization
constants of the axial-vector and vector currents, $Z_A$ and $Z_V$, 
as a function
of the heavy quark mass, $m_H$, can be found  in ref.~\cite{clv}.}.
In the following, we will denote these recipes with the generic name
of KLM prescriptions.
\par In analyzing our results we have been  particularly careful to isolate
the ground state to avoid higher-mass  state contamination.
We also studied the dependence of the final results on the
extrapolation in the heavy and light quark masses. A poor control
of these aspects in  the extraction of $f_P$ can mimic spurious
$O(a)$ effects. \par 
The study  of the effects which can fake discretization errors,
combined with the high statistics of the numerical runs, 
leads us to the  conclusion that a even higher statistics and
a larger spread of $a$ values are required to uncover
 satisfactorily the $O(a)$ dependence of the decay
constants.
Further studies at smaller values of the lattice spacing (corresponding to
$\beta=6.4$ and $6.6$) with comparable (or smaller) statistical
errors and similar (or larger) physical volumes
are required to reduce this source of uncertainty.
\par The main physical results of our study have been given in the abstract. 
They substantially  agree 
with recent compilations made in refs.~\cite{flynn} and \cite{beauty96}.
Preliminary results of this study can be found in ref.~\cite{oldall}.
\section{Details of the analysis}
\label{sec:anal}
In this section, we describe in detail the extraction of the  decay constants
from two-point correlation functions and  the extrapolation of the results
in the heavy and light quark masses to the physical points (including the
calibration of the lattice spacing). We discuss systematic errors
coming from the time interval of the fit, from the method used to derive
the decay constant, from the fit used for the extrapolation
in the heavy quark mass or in the light quark mass, 
and from the choice of the physical scale
(from the string tension $\sigma$, the mass of the rho mass  $M_\rho$ or from
$f_\pi$).
\subsection{Extraction of the decay constants}
\label{subsec:extra}
Consider the correlation function
\begin{equation}
C_{AP}(t) =
\sum_{\vec{x}} \langle 0 | A_0 (\vec{x},t) P^{\dag} (\vec{0},0) | 0 \rangle,
\label{eq:CSL}
\end{equation}
where $P(x) = \overline{Q}(x) \gamma_5 q(x)$.
We also introduce 
\begin{equation}
C_{PP}(t) =
\sum_{\vec{x}} \langle 0 | P (\vec{x},t) P^{\dag} (\vec{0},0) | 0 \rangle.
\label{eq:CSS}
\end{equation}
At large Euclidean times $C_{AP}(t)$ and $C_{PP}(t)$ behave as
\be C_{AP}(t) = \frac{Z_{AP}}{M_P} e^{-M_P T/2} \sinh(M_P(T/2-t))\, , 
\label{eq:sin}
\ee \be
C_{PP}(t) = \frac{Z_{PP}}{M_P} e^{-M_P T/2} \cosh(M_P(T/2-t)) \, , \label{eq:cos}
 \ee
where $T$ is the temporal extension of the lattice.
We extract the raw lattice value of $f_P$, $f_P^{latt}$,
using the usual ratio method
\begin{equation}
f_P^{latt} = 
\left< \frac{C_{AP}(t)}{C_{PP}(t)}
 \coth(M_P(T/2-t)) \right> \frac{\sqrt{Z_{PP}}}{M_P},
\label{eq:fplat}
\end{equation}
where $\langle...\rangle$ is a weighted average over a given time interval 
$t_1$--$t_2$.
$M_P$ and $Z_{PP}$ are extracted from a fit of $C_{PP}(t)$ as a function of 
$t$
to the expression given in eq.~(\ref{eq:cos}) in the same interval $t_1$--$t_2$
in which we compute $f_P^{latt}$.  The values of $t_1$--$t_2$ used to extract
all the results reported below are given, for the different simulations,
in table \ref{tab:parameters}. \par
The physical value of $f_P$ is then simply given by 
\begin{equation}
f_P \equiv \frac{\langle 0 | A_0 | P(\vec{p}=0) \rangle}{M_P}
      =     f_P^{latt} \; Z_{A}(a) \; a^{-1} \, .
\end{equation}
\par Alternatively, we can extract the decay constant of the heavier mesons,
by normalizing it to $f_\pi$ (or to $f_K$), 
defined as the pseudoscalar decay constant
computed in the chiral limit
\be f_P= R_P \times f_\pi^{exp}= \frac{f_P^{latt}(M_P)}{f_\pi^{latt}}
\times f_\pi^{exp} \, , \label{rp} \ee
where $f_\pi^{latt}=f_P^{latt}(M_P=0)$. In the following, it will be useful
to introduce also $R_{P_s}=f^{latt}_{P_s}/f^{latt}_K$, where 
$f^{latt}_K=f^{latt}_P(M_P=M_K)$ and $f_{P_s}^{latt}$ is $f_P^{latt}$ computed at
a value of the light quark mass corresponding to the strange quark,
$m_s$.
We have then 
\be f_{P_s}= R_{P_s} \times f_K^{exp} \, . \label{rps} \ee
\par We now discuss the choice of the time intervals given in table 
\ref{tab:parameters} and the size of the systematic error due to higher mass 
states. In all the cases, the range was chosen by demanding
that the contamination of the excited states in the parameters
of the fits (ie. $\langle\frac{C_{AP}}{C_{PP}} \coth\rangle$,
$M_P$ and $Z_{PP}$) is at most 20\% of the statistical error.
We impose this strict criteria to stop the systematic effects of the higher
states swamping the $a$ dependence of $f_P^{latt}$.
In particular, we have verified that,
due to the contamination of the excited states,
 the error  on  the decay constant
$f_P$, for values of the heavy and light quark masses
close to the appropriate ones for the $D_s$ meson, i.e. without almost any
 extrapolation in $m_H$ and $m_s$,
is of  about $2$~MeV (assuming $a^{-1}(\beta=6.0)=2$
GeV and $a^{-1}(\beta=6.2)=2.6$ GeV)   and that the relative error 
on $R_P$ is $\delta R_P/R_P \sim 0.005$.  This makes us confident that
 this error is much smaller than the other ones (the statistical error,
 the error  due 
to the extrapolation in the quark masses and the error coming from
the calibration of the lattice spacing).
\subsection{Extrapolation of the raw results} 
In order to obtain the physical values of $f_D$, $f_{D_s}$, etc.,
we have to extrapolate $f_P$ both in the heavy and light quark masses,
to fix the value of the lattice spacing in physical units and to determine
 $Z_A$ (or to use $R_P$ to extract the decay constant). \par
The best  method to monitor $O(a)$ effects  is to study $R_P$, computed for
a fixed physical value of the meson mass, using  the lattice spacing
$a$ calibrated by the string tension  for which, in the quenched approximation,
 errors are of $O(a^2)$.  By extracting the decay constant from $R_P$, 
we eliminate the apparent $a$-dependence coming from the variation of $Z_A$
with $\beta$. $Z_A$ is in fact independent of quark masses,
except for $O(a)$ corrections. Since  meson masses are not 
exactly the same for the different simulations, we are forced to make some
extrapolation both in the heavy and the light quark masses. In order to minimize
 the extrapolation, which will be discussed below, and at the
same time obtain  a physical quantity, we have chosen $R_{D_s}=f_{D_s}/f_K$,
which is obtained from $R_{P_s}$~\footnote{ It may also be that for this
quantity the error due to the quenched approximation is smaller, although
we have not checked this point yet.}.
Indeed,  for all the runs listed in table \ref{tab:parameters}, 
one of the heavy quark and one of the light
quark masses 
are very close to their physical values ($m_{charm}$ and $m_s$ respectively), 
and so in this case the effect of the
extrapolation is negligible.
Only the extrapolation
in one of the light quark masses (to the chiral limit)  is then relevant
in order to get $R_{D_s}$.
It turns out, see  table \ref{tab:rps},  that for $R_{D_s}$ the difference 
between the value obtained with a linear or a quadratic fit to $R_{P_s}$
is (at most) $2$ \%, corresponding to an error of about (less) $4$~MeV
for $f_{D_s}$.
This is to be contrasted with $R_D=f_D/f_\pi$ as obtained from a fit to $R_P$.
In this case, differences between  values obtained from a linear
or a  quadratic fit can be  as large as $10$ \%, mostly due to a quadratic 
dependence of $f_\pi$ on the light quark masses. 
The relevance of these differences for the physical results
will be discussed in the next section. \par 
As for the mass of the charmed and strange quarks, they have been determined
by fixing the  $D$- and $K$-meson masses to their physical values.
The choice of the physical quantities used to fix the different  quark masses
may also affect the effective $a$-dependence of the decay constants, since
the spectrum also suffers from discretization errors.  A
different possibility for the method used to fix $m_s$ 
will be discussed later on. 
\begin{table}
\begin{center}
\begin{tabular}{||cc|ccccc||}
\hline\hline         %123456
Method&&C60&C62&W60&W62a&W62b\\
\hline
&linear& 1.56(3)& 1.48(6)&1.11(3)&1.23(4)& 1.19(5)  \\
$R_{D_s}=$&quadratic &1.57(4)&1.49(7)&1.13(4)&1.25(5)&1.20(5)\\
$f_{D_s}/f_K$&linear KLM&1.59(3)&1.50(6)&1.48(6)&1.52(5)&1.47(6)\\
&quadratic KLM&1.61(4)&1.51(7)&1.51(5)&1.55(7)&1.47(6)\\
\hline \hline
&linear&1.63(4)&1.58(8)&1.14(4)&1.31(6)& 1.25(7)\\
$R_{D}=$&quadratic &1.69(8)&1.73(16)&1.19(8)&1.43(11)&1.28(9)\\
$f_{D}/f_\pi$&linear KLM&1.67(4)&1.60(8)&1.52(5)&1.61(7)&1.53(8)\\
&quadratic KLM&1.72(8)&1.75(16)&1.59(10)&1.77(13)&1.59(11)\\
\hline
\hline
$f_{D_s}/f_D$&linear&1.08(1)&1.07(2)&1.06(1)&1.07(1)&1.09(2)  \\ 
&quadratic &1.09(3)&1.04(4)&1.06(3)&1.06(3)&1.13(3)\\
\hline
\hline
\end{tabular}
\end{center}
\caption{\it{Summary of the physical results for $R_{D_s}=f_{D_s}/f_K$ and
$R_D=f_D/f_\pi$ obtained by extrapolating  $R_P$ and $R_{P_s}$.
We also give $f_{D_s}/f_D$. ``linear" and ``quadratic" refer to the
fit in the light quark masses.}}\label{tab:rps} 

\end{table}
Our results for $R_D$,
$R_{D_s}$ and $f_{D_s}/f_D$ are given in table \ref{tab:rps} for the following
cases:  i)~run C60, 
ii)~run C62, iii)~run C60 with KLM corrections, iv)~run C62 
with KLM corrections, v)~run W60, vi)~run W62 (a and b), vii)~run 
W60 with KLM corrections (KLM-Wilson), viii)~run W62 with KLM corrections
(a and b). \par
Since several  versions of the KLM prescription exist and, in the
Clover case, it must be implemented  from $O(a^2)$ only \cite{clv}, we now
state the recipe used in this work to obtain the results in table \ref{tab:rps}.
We corrected $f_P^{latt}$, as obtained from
eq.~(\ref{eq:fplat}), for any given
pair of values of the quark masses ($m_{1,2}$), by multiplying it by the factor
\be {\cal F}^W_{KLM}= \sqrt{(1+am_1)(1+am_2)} \, , \label{eq:wklm} \ee
in the Wilson case and by the factor \cite{clv}
\be {\cal F}^C_{KLM}= \frac{{\cal F}^W_{KLM}}{{F}_1{F}_2} \, , \label{eq:cklm} 
\ee
in the Clover case~\footnote{ The factor $\sqrt{4 K_1 K_2}$, where $K_{1,2}$
are the hopping parameters corresponding to the masses $m_{1,2}$, is always
included in our definition of the currents.}, where 
 \beqn {F}_{1,2} = 1+\frac{1}{4} \left[
(1+a m_{1,2}) - (1+a m_{1,2})^{-1} \right]  \, . \nonumber \eeqn 
In table \ref{tab:rps}, results from  linear and  quadratic fits
in the light quark masses  are both given. We do not give
results for $f_{D_s}/f_D$ with KLM corrections because they 
essentially cancel out for this ratio.
\par We have the following observations on the results given in table
\ref{tab:rps}. a) Without KLM factors the results in the Wilson case
are incompatible with those obtained with the Clover action. b) KLM-Wilson
 results are compatible 
 (indistinguishable) with  the Clover  results
at $\beta=6.0$ ($\beta=6.2$). c) Within  the statistical errors which are
 of the order of 3\%  
(corresponding to about  6~MeV of error for $f_{D_s}$),
KLM-Wilson results do not exhibit  any appreciable  $a$-dependence;
 d) Even with such small statistical
errors, it is impossible to decide whether there is a systematic shift
 of the Clover results between $\beta=6.0$
and $\beta=6.2$ or the difference only comes from a statistical fluctuation.  
 This is due to several reasons: differences between KLM-Wilson and Clover 
 results  at
$\beta=6.0$  are  of the same size than those between $\beta=6.0$ and
 $\beta=6.2$ in the Clover case; in the Clover case,   the KLM corrections
 to $f_{D_S}/f_K$ go in the wrong direction (they increase
the difference between the results at $\beta=6.0$ and $6.2$).
If we fit $R_P$ linearly in the light quark masses,
the KLM corrections increase
the difference between the results at $\beta=6.0$ and $6.2$
also in the case of  $R_D$. This is opposite to what happens 
with a  quadratic fit, although  within larger errors:
in this case the KLM corrections reduce the differences between the results
at $\beta=6.0$ and $6.2$. e) 
In the  Wilson  case, the variations observed 
between two different runs at $\beta=6.2$
 are about one half of the differences between the two Clover results
$\beta=6.0$ and $\beta=6.2$. 
 \par From the above considerations, it is clear that any attempt to extrapolate
our results to $a=0$ in order to reduce discretization error would be
fruitless.
This is even more true because, as observed in ref.~\cite{aggr},
 the results of the extrapolation are extremely sensitive to the  choice of
 the scale, given the small range in $a$ at disposal. We could have easily
 done a high statistics, very accurate  calculation at  lower 
 values of $\beta$, e.g. $\beta=5.7$, to increase the range in
 $a$. We believe, however, that at such large values of $a$ the  
 behaviour of the lattice dynamics is very different from the 
 continuum one, so that the inclusion of this point could fake
completely the final result. 
This is true not only for our data, which have tiny statistical errors, and
have been obtained  on (approximatively) the same physical volume, but also
for many other analyses which have been recently presented in the literature 
\cite{flynn}. With this  we do not mean that in the KLM-Wilson and Clover
cases there aren't
sizeable $O(a)$ effects for $m_H$ around the charm quark mass
\footnote{ It is evident that without the KLM factors 
these effects are very large in the Wilson case.} 
(indeed there are indications, from  calculations of the
dependence of the effective $Z_A$ on $m_H$,  that these effects  may be large
\cite{clv}). We want to stress that, in spite of the very good 
accuracy of our data,   it would be illusory to try to correct
$O(a)$ effects by extrapolating to $a=0$. This can be done only
by  enlarging the range of $a$ towards even smaller
values, corresponding to 
$\beta=6.4$ or even $6.6$, while keeping the physical volume, 
and the statistical accuracy the same as 
at $\beta=6.0$ and $6.2$, see also ref.~\cite{cracra}.
\subsection{Other sources of uncertainty}
In this subsection, we  discuss some tests which we performed in order
 to check the stability of the results for the D-meson decay constants
(see table \ref{tab:rps}) and the B-meson decay constants (presented in table
\ref{tab:rbs}).  These calculations consist in extracting the 
physical results
 with different assumptions on the parameters and the method chosen
 for the extrapolation. \par
 To be specific, we consider $R_{D_s}$, obtained in the Clover case
 from a linear fit in the light quark mass and (then) from  a fit in the
 heavy quark mass (at fixed $m_s$) of the form 
\be f_{P_s} \sqrt{M_{P_s}}
\approx \Phi^{inf}_s + \frac{\Phi_s'}{M_{P_s}} + \frac{\Phi_s''}{M_{P_s}^2} 
+ \dots 
\, , \label{eq:phi} \ee
and similarly for $f_P \sqrt{M_P}$.
 $\Phi^{inf}_s$, $\Phi_s'$, $\Phi''_s$ are functions which are expected to
 depend
logarithmically on $m_H$  but have been taken constant in the
fit. For $R_{D_s}$ the inclusion of the
quadratic term of eq.~(\ref{eq:phi}) in the fit
is immaterial, since it amounts to a difference of less
than about $3$~MeV for $f_{D_s}$.  In the $B$-case, the difference is of
about  $5$~MeV for $f_{B_s}$.
 Equation~(\ref{eq:phi}) has also been used to extrapolate in
$m_H$ the results of table \ref{tab:rps}, discussed in the previous
subsection.  If not stated otherwise, the fits in the heavy quark mass
always include the quadratic term.
\par We have considered the following cases:
\begin{enumerate} \item \underline{Choice of the scale.}  Although  $R_{D_s}$ is
a dimensionless quantity, the calibration  of the lattice spacing can affect
its value because it enters
in the determination of the values of the quark masses at which we extrapolate
$R_{P_s}$.  At $\beta=6.2$ with the Clover action, using a linear fit
in the light quark masses, we obtain $R_{D_s}=1.48(6)$ (scale from $\sigma$),
$1.48(6)$ (scale from $M_\rho$), $1.47(5)$ (scale from
$f_\pi$ with $Z_A=1.045$ as determined
non-perturbatively with the Ward identities \cite{ukqcdza}) and 
$R_{D_s}=1.47(6)$ (scale from $K^*$, using the lp-plane method of 
ref.~\cite{aggr}, see below). 
A similar result is obtained at $\beta=6.0$ where we used $Z_A=1.06$ 
\cite{mpsv}.
Thus the error due to the calibration of the lattice spacing is  negligible
for this ratio. 
\item \underline{Strange quark mass}: There are different equivalent methods to
fix $m_s$, from $M_K$, from $M_{K^*}$, or with the ``lattice physical plane"
(lp-plane) method
which has been used in refs.~\cite{aggr,lac} and that we briefly recall below.
All of these methods should give the same value of $m_s$, apart from
discretization errors and quenching effects. Thus we also computed 
$R_{D_s}$ by fixing $m_s$ with the lp-plane method for comparison.  With
this method, for the problem at hand, one defines a physical plane
[$M_Va$, $(M_Pa)^2$], where $M_V$ is the mass
of the vector meson. In  [$M_Va$, $(M_Pa)^2$], assuming that only linear
terms are important, one looks for the point where $M_V/M_P$ coincides with
the physical value $M_{K^*}/M_K$, and reads off the value 
of $m_s$ which is then used to evaluate  $R_{D_s}$. The results that we obtain
in this case are  indistinguishable from those given
in table \ref{tab:rps}. 
\item \underline{Extrapolation in the heavy quark mass}. Among the suggestions
introduced with the aim of  correcting 
$O(a)$ effects in the Wilson  case, it has been proposed \cite{BLS}, 
besides rescaling the quark fields  according
to the KLM prescription, to shift the mass $M_P$  by
\begin{equation}
M_Pa \to \tilde{M}_Pa = M_Pa - \tilde{m}_H + \overline{m}_H
 \, ,            \;\;\;\;\;\;
 (Q,q) \to {\cal N}(K_{H,l}) \; (Q,q)  \, ,
\label{eq:LM}
\end{equation}
where \beqn {\cal N}(K_H) &=& (\frac{4 K_{crit}}{K_H} - 3)^{1/2}\, ,
\;\;\;\;\;\;
 \tilde{m}_H = log({\cal N}^2(K_H)) \, , \nonumber \\
\overline{m}_H &=& e^{\tilde{m}_H} \sinh \; \tilde{m}_H / (\sinh \; \tilde{m}_H 
+ 1)\, .  \eeqn
$K_{crit}$ is the critical value of the Wilson hopping parameter and 
${\cal N}(K_H)$ is simply a different form  of the KLM 
prescription which was used in eq.(\ref{eq:wklm}).
We have then fitted the result in $\tilde{M}_P$ instead than 
$M_P$. The values of  $R_{D_s}$ change as follows:
$R_{D_s}=1.48(6) \to 1.44(3)$ (W60), $1.52(5) \to 1.51(5)$ (W62a)
and $1.47(6) \to 1.47(6)$ (W62b). Thus, the results
obtained by using the recipe in eq.(\ref{eq:LM}) are indistinguishable, within
the errors from the KLM-Wilson results reported in table \ref{tab:rps}.
This is not surprising because $\tilde{M}_P$ is only shifted by a small
amount from $M_P$ and the range of the extrapolation is very small. 
The effect is only  slightly larger after the extrapolation
to  $B$-mesons, and give a small shift, corresponding to an uncertainty
of about $5$~MeV for $f_{B_s}$. We included this error, by combining it
in quadrature with other ones, in the final evaluation of 
$f_{B_s}$ and $f_{B}$, which can be found in  subsection \ref{subsec:fb}.
\end{enumerate}
This concludes the discussion of several  minor uncertainties
in the calculation of $R_{D_s}$, $f_{D_s}/f_D$, $R_{B_s}$ and
 $f_{B_s}/f_B$. 
All of them have little influence in both the $D$- and $B$-meson cases.
\section{Physical results}
On the basis of the discussion of section \ref{sec:anal}, we are now
ready to present our final results and errors. We will first give
the results for charmed mesons, for which the extrapolation in the heavy
quark mass is not a relevant  source of systematic uncertainties, and
then discuss the $B$-meson case.
\subsection{Results for $f_{D_s}$ and $f_D$}
\label{subsec:fd}
From section  \ref{sec:anal}, we learned that an extrapolation to $a=0$
it is not possible at this stage. Thus we believe that the best estimate
of $f_{D_s}$ is obtained from the Clover data at $\beta=6.2$, by using
$R_{D_s}$ in eq.~(\ref{rps}) (from a linear fit in the light quark masses,
a quadratic fit in $1/M_{P_s}$ and without any KLM factor).
As for the error, we take as a conservative
estimate of the  discretization error the difference between the results
 obtained
at $\beta=6.0$ and $6.2$, and combine it in quadrature with the statistical
one. This gives $R_{D_s}=1.48(10)$ from which, by using $f_K^{\exp}=
159.8$~MeV, we obtain $f_{D_s}=237 \pm 16$~MeV.  By using $R_{D_s}=1.48(10)$
combined with $f_{D_s}/f_D=1.07(4)$ we obtain $f_D=221 \pm 17$~MeV.
We believe that these results, which have also been given in the abstract,
are our ``best" results. They are in very good
agreement with the compilation of lattice calculations
presented in refs.~\cite{flynn,beauty96}.
Since at $\beta=6.2$ the ratio $R_{D_s}$ is essentially identical
in the Clover and KLM-Wilson case,  
the Wilson results do not add much information, besides
giving  a consistency check.
\par If we extract $f_D$ from $R_D$ by using eq.~(\ref{rp}) instead,
we have to take into account the larger differences due to the use of the
linear or the quadratic extrapolation in the light quark masses. We also
take into account an error on   $f_D$ of about $6$~MeV, which comes from
the difference in the results obtained with a linear or quadratic fit of $f_P$
to eq.~(\ref{eq:phi}).  
 Proceeding as before, i.e.  taking  as 
 discretization error the difference between the result obtained
at $\beta=6.0$ and $6.2$, and combining it in quadrature with the statistical
error, we find $f_D=209 \pm 16$~MeV (linear) or $f_D=228 \pm 22$~MeV 
(quadratic). 
Our previous result for $f_D$ sits in the middle of these two numbers.
\par To reduce the uncertainties due to the extrapolation in the light
quark masses, and which are mainly related to the quadratic dependence 
of $f_\pi$ on the quark masses, we can compute directly
$f_D$, by assuming  some value
for the calibration of $a$ and for   $Z_A$.
By taking $a^{-1}$ as determined from $M_\rho$, which in the Clover
case is $a^{-1}(\beta=6.0)=1.92(11)$ and $a^{-1}(\beta=6.2)=2.56(21)$,
$Z_A(\beta=6.0)=1.06$ and $Z_A(\beta=6.2)=1.045$ 
\cite{ukqcdza,mpsv}, and fixing
$m_s$ from $M_K$, we obtain (only linear fits in the light quark masses)
$f_D(\beta=6.0)=212\pm 20$~MeV, $f_D(\beta=6.2)=204\pm 20$~MeV,
 $f_{D_s}(\beta=6.0)=219\pm 17$~MeV
and $f_D(\beta=6.2)=230\pm 9$~MeV, which are slightly lower 
than the previous results, but perfectly consistent with them.
\par The  latter two determinations of $f_{D_s}$ and $f_D$ are, however,
subject to larger systematic effects,  either due to the choice of the fit or
to the assumptions for the values of $a$ and $Z_A$,  than those extracted
from $R_{D_s}$ and $f_{D_s}/f_D$. We therefore prefer these as our best values.
\subsection{Results for $f_{B_s}$ and $f_B$}
\label{subsec:fb}
\begin{table}
\begin{center}
\begin{tabular}{||cc|ccccc||}
\hline\hline         %123456
Method&                    &  C60  &  C62  &   W60 &  W62a &W62b  \\
\hline
            &linear        &1.48(7) &1.28(9) &0.79(4)&0.83(4)&0.84(5)\\
$f_{B_s}/f_K$&quadratic    &1.49(9) &1.27(12)&0.81(5)&0.84(5)&0.84(5)\\
             &linear KLM   &1.56(7) &1.33(9) &1.29(4)&1.26(6)&1.16(7)\\
             &quadratic KLM&1.57(12)&1.33(11)&1.33(6)&1.27(8)&1.16(7)\\
\hline \hline
             &linear       &1.53(9) &1.32(13)&0.81(5) &0.86(6) &0.86(6)\\
$f_{B}/f_\pi$&quadratic    &1.57(16)&1.37(66)&0.87(9) &0.91(11)&0.86(9)\\
             &linear KLM   &1.61(10)&1.38(13)&1.32(6) &1.31(8) &1.19(8)\\
             &quadratic KLM&1.65(32)&1.43(50)&1.41(11)&1.39(15)&1.19(11)\\
\hline
\hline
$f_{B_s}/f_B$&linear       &1.10(3)&1.14(6)&1.05(2)&1.10(3)&1.12(3)\\ 
             &quadratic    &1.13(7)&1.17(17) &1.03(6)&1.14(6)&1.20(7)\\
\hline
\hline
\end{tabular}
\end{center}
\caption{\it{Summary of the physical results for $f_{B_s}/f_K$,
$f_B/f_\pi$ obtained by extrapolating  $R_P$ and $R_{P_s}$.
We also give $f_{B_s}/f_B$. ``linear" and ``quadratic" refer to the
fit in the light quark masses.}}
\label{tab:rbs} 
\end{table}
In order to obtain $f_{B_s}$ and $f_B$, an extrapolation in the heavy quark
 mass
well outside the range available in our numerical  simulations is
necessary. Discretization errors can affect the final results in two ways.
Not only do they change the actual values of the decay constants, but also
deform the  dependence of $f_P$ on $m_H$. Moreover, points
obtained at the largest values of $m_H a$ become the most important, since
we extrapolate in the direction of even larger values of $m_H$.
\par
\begin{figure}
\vspace{9pt}
\begin{center}\setlength{\unitlength}{1mm}
\begin{picture}(160,80)
\put(45,0){\epsfbox{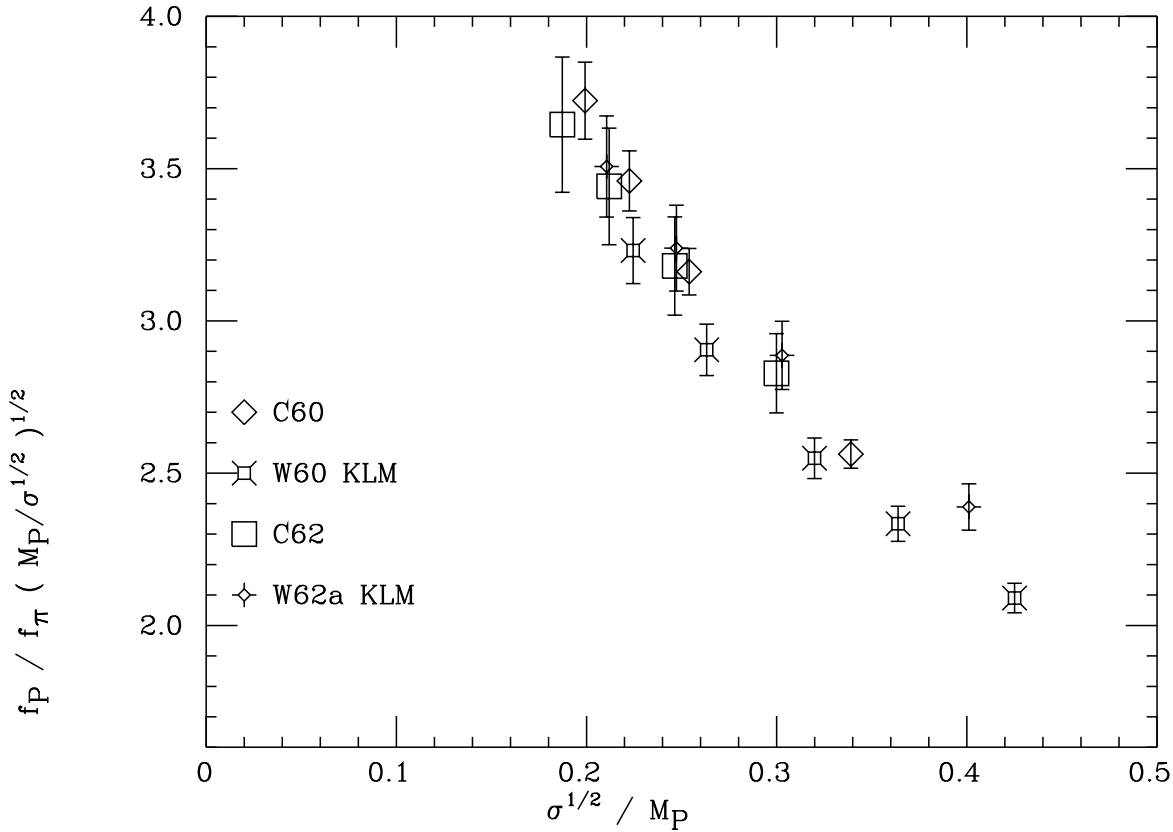}}
\end{picture}
\end{center}
\caption{\it{Dependence of 
$f_P/f_{\pi} (M_P/\sigma^{1/2})^{1/2}$ on $\sigma^{1/2}/M_P$.
For these points a linear extrapolation in the light quark masses
to the chiral limit has been used.}}
\label{fig:one}
\end{figure}

 In fig. \ref{fig:one}, we show the KLM-Wilson and Clover results
for $f_P/f_\pi \sqrt{M_P/\sigma^{1/2}}$ as a function of the dimensionless
scale $\sigma^{1/2}/M_P$. We note the remarkable
agreement between the scaled KLM-Wilson and the Clover data
(as was first observed in \cite{ukqcd}). In the figure,
we do not show the Wilson results
without the KLM prescription, because they are inconsistent between each other
(at the two values of $\beta$) and with the Clover results. \par
The results reported in table \ref{tab:rbs} were  obtained by  fitting
$R_{P_s}$ ($R_P$) to eq.~(\ref{eq:phi}), including the quadratic term.
We notice that all the differences observed in table \ref{tab:rps} for
$D$-mesons are amplified when we extrapolate to $B$-mesons.
Not only is this true for the results of the runs C60 and C62, where we
can suspect that the differences are due to discretization errors, but also 
for the runs W62a and W62b, performed at the same value of $\beta$.
Thus, the same general considerations made in section \ref{sec:anal}
for $R_{D_s}$ apply here. In order to extract our ``best" values
for $f_{B_s}$ and $f_{B}$ we proceed exactly as in subsection
\ref{subsec:fd}.  We obtain $f_{B_s}=205 \pm 15 \pm 31$~MeV=
$205 \pm 35$~MeV, where the second error ($31$~MeV) is the ``discretization"
error, estimated by comparing the results from C60 and C62.
We also obtain $f_B = 180 \pm 32$~MeV.
As in the $D$-meson case, these results are in good 
agreement with the compilation of lattice calculations
presented in refs.~\cite{flynn,beauty96}.
%___________________________________________________________________________
%___________________________________________________________________________
\section{Conclusion}
Using several runs obtained using the Wilson and
Clover actions at $\beta=6.0$ and $6.2$,
from a careful analysis of all possible effects which can
fake discretization errors, we conclude
that, using the methods outlined above, we require even higher statistics,
or a larger spread of $a$ values,
to uncover satisfactorily the $O(a)$ dependence of the decay
constants. We believe that other studies of the same problem have
the same difficulties as us in controlling and correcting discretization
errors. Further studies, with comparable (or smaller) statistical
errors and physical volume,
 at smaller values of the lattice spacing,  corresponding to
 $\beta=6.4$ and $6.6$, are required to reduce this source of uncertainty.
The use of the action of ref.~\cite{luescher} can be of great help  in this
respect. 
\par By assuming quite conservative discretization errors we found
\beqn f_{D_s}&=&237 \pm 16 \  \mbox{MeV} \, , \quad
f_{D} \; = \; 221 \pm 17 \  \mbox{MeV} \, , \nonumber \\ 
f_{B_s}&=&205 \pm 35 \  \mbox{MeV} \, , \quad
f_{B} \; = \; 180 \pm 32 \  \mbox{MeV} \, ,
\nonumber \\ \frac{f_{D_s}}{f_{D}}&=& 1.07 \pm 0.04
 \, , \quad  \frac{f_{B_s}}{f_{B}}= 1.14 \pm 0.08 \eeqn
in good agreement with previous estimates \cite{flynn,beauty96}.

\end{document}